\documentclass[aps,prl,twocolumn,showpacs,amsmath,amssymb,superscriptaddress]{revtex4}
\usepackage{graphicx}
\usepackage{amssymb}
\usepackage{natbib}
\usepackage{color}

\newcommand{\beq}{\begin{eqnarray}}
\newcommand{\eeq}{\end{eqnarray}}

\begin{document}
\title{Intertwined magnetic and nematic orders in semiconducting KFe$_{0.8}$Ag$_{1.2}$Te$_2$}
\author{Yu Song}
\email{yusong@berkeley.edu}
\affiliation{Department of Physics, University of California, Berkeley, California 94720, USA}
\affiliation{Materials Sciences Division, Lawrence Berkeley National Laboratory, Berkeley, California 94720, USA}
\author{Huibo Cao}
\affiliation{Neutron Scattering Division, Oak Ridge National Laboratory, Oak Ridge, Tennessee 37831, USA}
\author{B. C. Chakoumakos}
\affiliation{Neutron Scattering Division, Oak Ridge National Laboratory, Oak Ridge, Tennessee 37831, USA}
\author{Yang Zhao}
\affiliation{NIST Center for Neutron Research, National Institute of Standards and Technology, Gaithersburg, Maryland 20899, USA}
\affiliation{Department of Materials Science and Engineering, University of Maryland, College Park, Maryland 20742, USA}
\author{Aifeng Wang}
\affiliation{Condensed Matter Physics and Materials Science Department, Brookhaven National Laboratory, Upton, New York 11973, USA}
\author{Hechang Lei\footnote{Present address: Department of Physics, Renmin University, Beijing 100872, China.}}
\affiliation{Condensed Matter Physics and Materials Science Department, Brookhaven National Laboratory, Upton, New York 11973, USA}
\author{C. Petrovic}
\affiliation{Condensed Matter Physics and Materials Science Department, Brookhaven National Laboratory, Upton, New York 11973, USA}
\author{Robert J. Birgeneau}
\affiliation{Department of Physics, University of California, Berkeley, California 94720, USA}
\affiliation{Materials Sciences Division, Lawrence Berkeley National Laboratory, Berkeley, California 94720, USA}
\affiliation{Department of Materials Science and Engineering, University of California, Berkeley, California 94720, USA}

\begin{abstract}
Superconductivity in the iron pnictides emerges from metallic parent compounds exhibiting intertwined stripe-type magnetic order and nematic order, with itinerant electrons suggested to be essential for both. Here we use X-ray and neutron scattering to show that a similar intertwined state is realized in semiconducting KFe$_{0.8}$Ag$_{1.2}$Te$_2$ (K$_5$Fe$_4$Ag$_6$Te$_{10}$) without itinerant electrons. 
We find Fe atoms in KFe$_{0.8}$Ag$_{1.2}$Te$_2$ form isolated $2\times2$ blocks, separated by nonmagnetic Ag atoms.
Long-range magnetic order 
sets in below $T_{\rm N}\approx35$ K, with magnetic moments within the $2\times2$
Fe blocks ordering into the stripe-type configuration. A nematic order accompanies the magnetic transition, manifest as a structural distortion that breaks the fourfold rotational symmetry of the lattice. The nematic orders in KFe$_{0.8}$Ag$_{1.2}$Te$_2$ and iron pnictide parent compounds are similar in magnitude and how they relate to the magnetic order, indicating a common origin.    
Since KFe$_{0.8}$Ag$_{1.2}$Te$_2$ is a semiconductor without itinerant electrons, this indicates that local-moment magnetic interactions are integral to its magnetic and nematic orders, and such interactions may play a key role in iron-based superconductivity. 
  
\end{abstract}

\pacs{74.25.Ha, 74.70.-b, 78.70.Nx}

\maketitle

The parent compounds of the iron pnictide superconductors exhibit stripe-type magnetic order, typically accompanied or preceded by the onset of a nematic order that drives a concomitant tetragonal-to-orthorhombic structural transition \cite{GStewart_RMP,PDai_RMP,RMFernandes_NP,QMSi_NRM}. The onset of nematic order results in the differentiation of lattice spacings along the Fe-Fe bond directions, characterized by an orthorhombicity [$\delta = (a-b)/(a+b)$] of a few tenths of a percent. In the magnetically ordered state, spins are antiferromagnetically aligned along the longer axis and ferromagnetically aligned along the shorter axis \cite{Clarina,QHuang}. 

Although a multitude of different magnetic orders have been uncovered in related materials \cite{WBao2009,YSingh2009,DGQuirinale2013,XTan2016,WBao2011,FYe2011,MWang2011,BCSales2011,JZhao2012,HBCao2012,AFMay2012,MWang2015,KMTaddei2015,JMCaron2011,FDu2012,YSong2016,WWang2017}, observations of intertwined stripe-type magnetic order and nematic order are so far limited to iron pnictides such as LaFeAsO and BaFe$_2$As$_2$ \cite{PDai_RMP,Clarina,QHuang}. Upon doping, these intertwined orders are suppressed, giving way to superconductivity near the corresponding quantum critical points \cite{TShibauchi2014}, evidencing their intimate roles in iron-based superconductivity. The ubiquity of stripe-type magnetic \cite{ADChristianson2008,QWang2016} and nematic \cite{HHKuo2016} fluctuations in iron-based superconductors further reinforces this view, setting intertwined magnetic and nematic orders at the center of iron-based superconductivity research.

The stripe-type magnetic order has been suggested to originate from nesting of itinerant electrons \cite{IIMazin2010,PHirschfeld}, or local-moments on the verge of a Mott-insulating state \cite{QMSi2008}. For the nematic order, while a ferroelastic origin has been ruled out \cite{JHChu1,JHChu2}, the question whether it results from the magnetic or the orbital degree of freedom remains open \cite{RMFernandes_NP}. In the magnetic scenario, nematic order results from the breaking of fourfold rotational symmetry in the spin-spin correlations of stripe-type magnetism \cite{CFang2008,CXu2008,RMFernandes2012_1,RMFernandes2012_2,SLiang2013}. In the orbital picture, the onset of nematic order is due to the lifting of orbital degeneracy \cite{WLv2009,CCLee2009}, manifest as the splitting of bands with $d_{xz}$ and $d_{yz}$ characters near the Fermi level \cite{MYi2011}. In both cases, itinerant electrons near the Fermi level have been suggested to be essential for the nematic state \cite{RMFernandes2012_1,RMFernandes2012_2,MYi2011}. Alternatively, the nematic state can result solely from local-moment stripe-type magnetism, without the need of itinerant electrons \cite{CFang2008,CXu2008}.

In this work we reveal the presence of intertwined stripe-type magnetic order and nematic order in semiconducting KFe$_{0.8}$Ag$_{1.2}$Te$_2$ single crystals using X-ray and neutron diffraction. We find that the Fe atoms in KFe$_{0.8}$Ag$_{1.2}$Te$_2$ order into 2$\times$2 blocks separated by non-magnetic Ag atoms, forming a body-centered $\sqrt{5}\times\sqrt{5}$ tetragonal superstructure with the stoichiometric composition K$_5$Fe$_4$Ag$_6$Te$_{10}$ [Fig. 1(a) and (d)]. Below $T_{\rm N}$, spins within each block order into a collinear stripe-type configuration, while the orientation of the spins is modulated from block to block by an incommensurate propagation vector [Fig. 1(b)]. A structural transition reflecting the onset of a nematic order accompanies the magnetic transition, breaking the fourfold rotational symmetry of the lattice. The structural transition results in an expanded lattice spacing along the Fe-Fe direction with antiferromagnetically aligned spins, and a contracted lattice spacing along the Fe-Fe direction with ferromagnetically aligned spins [Fig. 1(b)], similar to the parent compounds of the iron pnictides. In addition, the magnitude of the distortion is also similar to those in the iron pnictides, pointing to a common origin of the intertwined orders. Importantly, for KFe$_{0.8}$Ag$_{1.2}$Te$_2$ the essential physics is already present in isolated 2$\times$2 Fe blocks, which builds up the Fe-pnictogen/chalcogen planes of iron-based superconductors. Because KFe$_{0.8}$Ag$_{1.2}$Te$_2$ is a semiconductor without itinerant electrons, its intertwined magnetic and nematic orders likely result from local-moment magnetism, with the underlying magnetic interactions also important for iron-based superconductivity. 

\begin{figure}[t]
	\includegraphics[scale=.50]{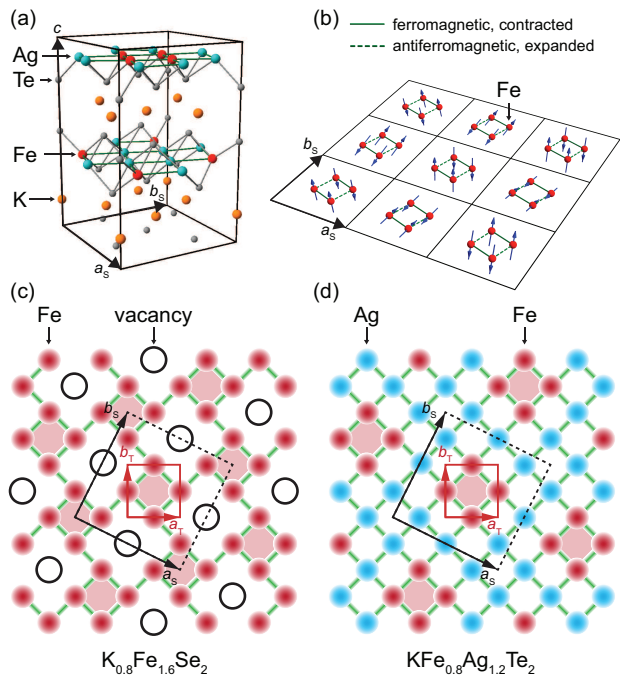}
	\caption{
		(Color online) (a) Crystal structure of KFe$_{0.8}$Ag$_{1.2}$Te$_2$, with the unit cell expanded by $\sqrt{5}\times\sqrt{5}$ in the $ab$ plane relative to the $I4$/$mmm$ unit cell. (b) Magnetic structure of KFe$_{0.8}$Ag$_{1.2}$Te$_2$, with only Fe atoms shown. (c) Schematic of Fe-plane in K$_{0.8}$Fe$_{1.6}$Se$_2$. (d) Schematic of Fe-Ag plane in KFe$_{0.8}$Ag$_{1.2}$Te$_2$. The solid-line boxes in (c) and (d) represent the $I4$/$mmm$ unit cells, and the dashed-line boxes are the $\sqrt{5}\times\sqrt{5}$ superstructure unit cells. The squares shaded in red highlight the $2\times2$ Fe blocks.  
	}
\end{figure}

The growth and physical properties of KFe$_{0.8}$Ag$_{1.2}$Te$_2$ single crystals have been described previously \cite{HLei2011}. Neutron scattering measurements were carried out on the four circle diffractometer HB-3A, HFIR, Oak Ridge National Laboratory, and the BT-4 triple-axis spectrometer at the NIST Center for Neutron Research (NCNR). Single-crystal X-ray diffraction data were collected at 260 K using a Rigaku XtaLAB PRO diffractometer. Experimental details are described in the Supplemental Material \cite{SI}. 

\begin{figure}[t]
	\includegraphics[scale=.47]{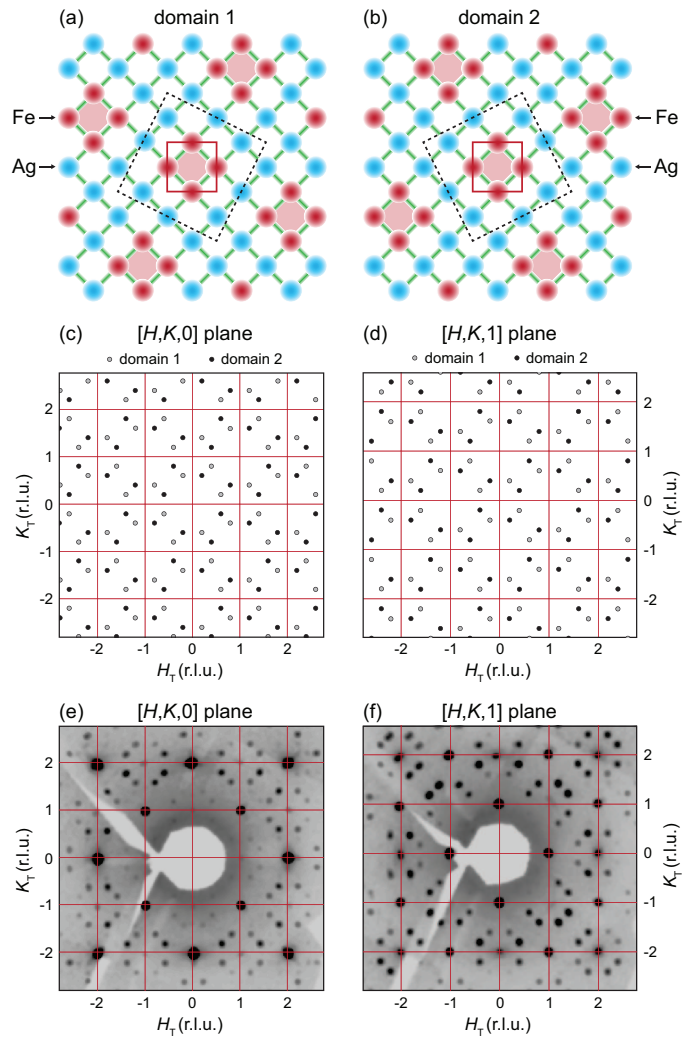}
	\caption{ 
		(Color online) Ordering of Fe and Ag for superstructure (a) domain 1 and (b) domain 2. The presence of both domains results in superstructure peaks in reciprocal space as shown for the (c) $[H,K,0]$ and (d) $[H,K,1]$ planes. The superstructure peaks due to the two domains are well separated. X-ray diffraction data of KFe$_{0.8}$Ag$_{1.2}$Te$_2$ in the (e) $[H,K,0]$ and (f) $[H,K,1]$ scattering planes measured at 260 K. The red horizontal and vertical lines represent the reciprocal lattice of the $I4$/$mmm$ unit cell.
	}
\end{figure}

Previous X-ray powder diffraction measurements suggested that KFe$_{0.8}$Ag$_{1.2}$Te$_2$ is isostructural to tetragonal BaFe$_2$As$_2$ (space group $I4$/$mmm$, $a_{\rm T}=b_{\rm T}\approx$ 4.37 {\AA} and $c\approx$ 14.95 {\AA}), with disordered Fe and Ag \cite{HLei2011}. Our single crystal X-ray diffraction measurements reveal a series of superstructure peaks [Figs. 2(c)-(f)] that indicate a unit cell expanded by $\sqrt{5}\times\sqrt{5}$ in the $ab$ plane due to the ordering of Fe and Ag, forming two structural domains that are mirror images of each other [Figs. 2(a), (b)], similar to iron-vacancy-ordered K$_{0.8}$Fe$_{1.6}$Se$_2$ (K$_2$Fe$_4$Se$_5$) [Figs. 1(c), (d)] \cite{WBao2011}. Reciprocal lattice vectors for the two superstructure domains can be obtained from that of the $I4$/$mmm$ unit cell in reciprocal lattice units (r. l. u.), using $H_{\rm S}=2H_{\rm T}-K_{\rm T}$, $K_{\rm S}=H_{\rm T}+2K_{\rm T}$ (the subscripts are used to distinguish vectors in different notations) for one structural domain [Fig. 2(a)], and $H_{\rm S}=2H_{\rm T}+K_{\rm T}$, $K_{\rm S}=-H_{\rm T}+2K_{\rm T}$ for the other [Fig. 2(b)]. The crystal structure of KFe$_{0.8}$Ag$_{1.2}$Te$_2$ is body-centered and exhibits four-fold rotational symmetry with space group $I4$, detailed in Table I. Bragg peaks associated with the superstructure are well separated in reciprocal space for the two structural domains, allowing them to be easily distinguished [Figs. 2(c), (d)]. 

\begin{table*}
	\caption{Refined structural parameters for KFe$_{0.8}$Ag$_{1.2}$Te$_2$ from X-ray diffraction data measured at 260 K. Space group is $I4$ with $a=b=9.7857(7)$ {\AA} and $c=14.933(3)$ \AA.}
	\begin{ruledtabular}
		\begin{tabular}{ccccccc}
			Atom &Site &$x$&$y$
			&$z$ &Occupancy&$U_{\rm eq}$ ({\AA}$^2$)\\
			\hline
			K& $2a$ & 1/2 & 1/2 &0.6472(9)
			& 1 & 0.043(2)  \\
			K& $8c$ & 0.5978(4) & 0.8044(4) &0.1680(7)
			& 1 & 0.0427(12) \\
			Fe& $8c$ &0.39925(18) & 0.69461(18) &0.9123(4)
			& 1 & 0.0196(4) \\
			Ag& $4b$ & 1/2 & 0 &0.91705(16)
			& 1 & 0.0324(6) \\
			Ag& $8c$ & 0.08740(16) & 0.79877(12) &0.91032(16)
			& 1 & 0.0335(5) \\
			Te& $2a$ & 0 & 0 &0.7851(2)
			& 1 & 0.0355(9) \\
			Te& $2a$ & 1/2 & 1/2 &0.01838(18)
			& 1 & 0.0281(6) \\
			Te& $8c$ & 0.30304(10) & 0.87686(17) &0.02736(10)
			&  1 & 0.0333(4) \\
			Te& $8c$ & 0.59148(10) & 0.78865(11) &0.79997(8)
			& 1 & 0.0316(4) \\
		\end{tabular}
	\end{ruledtabular}

\end{table*}

To elucidate the magnetic structure of KFe$_{0.8}$Ag$_{1.2}$Te$_2$, we systematically studied magnetic Bragg peaks exclusively associated with one of the superstructural domains [Fig. 2(a)] using HB-3A. We found that magnetic Bragg peaks occur at ${\bf Q}=\boldsymbol{\tau}_{\rm S}\pm{\bf q}_{\rm m}$, where $\boldsymbol{\tau}_{\rm S}$ are structural Bragg peaks of the $\sqrt{5}\times\sqrt{5}$ unit cell, and ${\bf q}_{\rm m}=(0.29(1),-0.26(1),0)_{\rm S}$ is the incommensurate magnetic propagation vector, suggesting that the magnetic structure does not exhibit four-fold rotational symmetry. Measurement of the magnetic intensity which scales like the magnetic order parameter squared, is shown in Fig. 3(a) for ${\bf Q}=(1,1,0)_{\rm S}+{\bf q}_{\rm m}$, revealing a clear onset of magnetic order below $T_{\rm N}\approx35$ K, in good agreement with susceptibility measurements \cite{HLei2011}. 


\begin{figure}[t]
	\includegraphics[scale=.47]{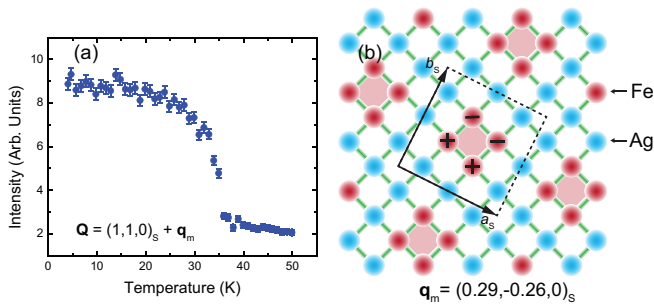}
	\caption{
		(Color online) (a) Temperature dependence of magnetic intensity at ${\bf Q}=(1,1,0)_{\rm S}+{\bf q}_{\rm m}=(1.29,0.74,0)_{\rm S}$. 
		(b) Schematic of stripe-type pattern within the $2\times2$ block associated with ${\bf q}_{\rm m}=(0.29,-0.26,0)_{\rm S}$.
		Error bars represent one standard deviation. 
	}
\end{figure}

We find the 
magnetic moments inside each Fe block form a collinear stripe-type pattern with an ordered moment of 2.11(3) $\mu_{\rm B}$/Fe, as shown in Fig. 1(b) and Fig. 3(b). 
The spin orientations of the $2\times2$ blocks are modulated from block to block by ${\bf q}_{\rm m}$, resulting from the weak spin anisotropy as indicated in susceptibility measurements \cite{HLei2011}, and the presence of small inter-block couplings.  
The magnetic unit cell of KFe$_{0.8}$Ag$_{1.2}$Te$_2$ contains a single $2\times2$ Fe block with 4 Fe atoms. This means that, compared to the magnetic structure of the iron pnictides which contains a single Fe atom, a magnetic structure factor $|F({\bf Q})|^2=|\sum_{j=1}^{4}{{\bf m}_j\exp{(-i{\bf Q}\cdot{\bf r}_j})}|^2$ modulates the intensities of the magnetic Bragg peaks that occur at ${\bf Q}=\boldsymbol{\tau}_{\rm S}\pm{\bf q}_{\rm m}$, and such a modulation provides direct evidence for the stripe-type configuration within the Fe blocks. 
While magnetic Bragg peaks in KFe$_{0.8}$Ag$_{1.2}$Te$_2$ occur at an incommensurate propagation vector associated with the $\sqrt{5}\times\sqrt{5}$ unit cell rather than the stripe vector ${\bf Q}=(0.5,0.5)_{\rm T}$ in iron pnictides, the magnetic structure factor $|F({\bf Q})|^2$ that modulates the intensities of the magnetic Bragg peaks is peaked at ${\bf Q}=(0.5,0.5)_{\rm T}$ positions, due to the stripe-type pattern within the Fe blocks (See Supplemental Material for details \cite{SI}).

Since the magnetic structure breaks fourfold rotational symmetry of the paramagnetic tetragonal lattice, it is important to clarify if a nematic order breaking the same symmetry is also present and how it relates to the nematic order in the iron pnictides. In single crystals of the iron pnictides such as BaFe$_2$As$_2$ \cite{MATanatar2009}, four orthorhombic domains form below the tetragonal-to-orthorhombic structural transition temperature [Figs. 4(a) and (c)], resulting in the splitting of (2,2,0)$_{\rm T}$ and (2,0,0)$_{\rm T}$ peaks. $(2,2,0)_{\rm T}$ splits into a quartet [Figs. 4(b) and (d)], with the separation of the peaks along the longitudinal and the transverse directions $dQ$ related to the orthorhombicity $\delta$ through $dQ\approx2\delta |{\bf Q}|$. While $(2,0,0)_{\rm T}$ also splits into four peaks, the splitting is mostly along the transverse direction with two peaks overlapping in the center straddled by the other two peaks on each side [Figs. 4(b) and (d)]; the separation between the two peaks on the sides is $dQ\approx4\delta|{\bf Q}|$.

Using BT-4 we carried out longitudinal and transverse scans at ${\bf Q}=(2,2,0)_{\rm T}$ and ${\bf Q}=(2,0,0)_{\rm T}$ below and above $T_{\rm N}$ in KFe$_{0.8}$Ag$_{1.2}$Te$_2$, to see if a similar distortion occurs. Clear broadening from 40 K to 6 K can be seen in the longitudinal and transverse scans at $(2,2,0)_{\rm T}$ and the transverse scan at $(2,0,0)_{\rm T}$, while a much smaller or no broadening is seen for the longitudinal scan at $(2,0,0)_{\rm T}$ [insets in Figs. 4(e)-(h)]. Using 
scans well above $T_{\rm N}$ ($T=70$ K) as peak line-shapes, we fitted longitudinal and transverse scans at $(2,2,0)_{\rm T}$ and longitudinal scans at $(2,0,0)_{\rm T}$ at different temperatures using two split peaks separated by $dQ$, and transverse scans at $(2,0,0)_{\rm T}$ using three peaks with the two side peaks separated by $dQ$. The fit results, after dividing by the corresponding $|{\bf Q}|$, are shown in Figs. 4(e)-(h). For $(2,2,0)_{\rm T}$, the splittings are similar along the longitudinal and transverse directions, whereas for $(2,0,0)_{\rm T}$ the splitting occurs predominantly along the transverse direction, demonstrating that the structural distortion is dominated by differing lattice spacings along the Fe-Fe bond directions, similar to iron pnictides such as BaFe$_2$As$_2$. Coupled with the $\sqrt{5}\times\sqrt{5}$ superstructure in KFe$_{0.8}$Ag$_{1.2}$Te$_2$, the crystal system would become monoclinic or triclinic since the angle between $a_{\rm S}$ and $b_{\rm S}$ is no longer 90$^\circ$; nonetheless, the main effect of the structural distortion is the differentiation of lattice spacings along the Fe-Fe direction that can be characterized by a similar orthorhombicity as defined for iron pnictides. For longitudinal and transverse scans at $(2,2,0)_{\rm T}$, $dQ/|{\bf Q}|\approx2\delta$; for the transverse scan at $(2,0,0)_{\rm T}$, $dQ/|{\bf Q}|\approx4\delta$. The values of $\delta$ obtained from Figs. 4(e)-(g) are consistent and average to $\delta\approx3.8\times10^{-3}$, which is close to $\delta\approx4.0\times10^{-3}$ for BaFe$_2$As$_2$ \cite{QHuang}. Moreover, we find that the expanded lattice spacing is associated with the antiferromagnetically aligned Fe-Fe bond direction, whereas the contracted lattice spacing is associated with the ferromagnetically aligned Fe-Fe bond direction (See Supplemental Material for details \cite{SI}), which is the same as in the iron pnictide parent compounds.

\begin{figure}[t]
	\includegraphics[scale=.47]{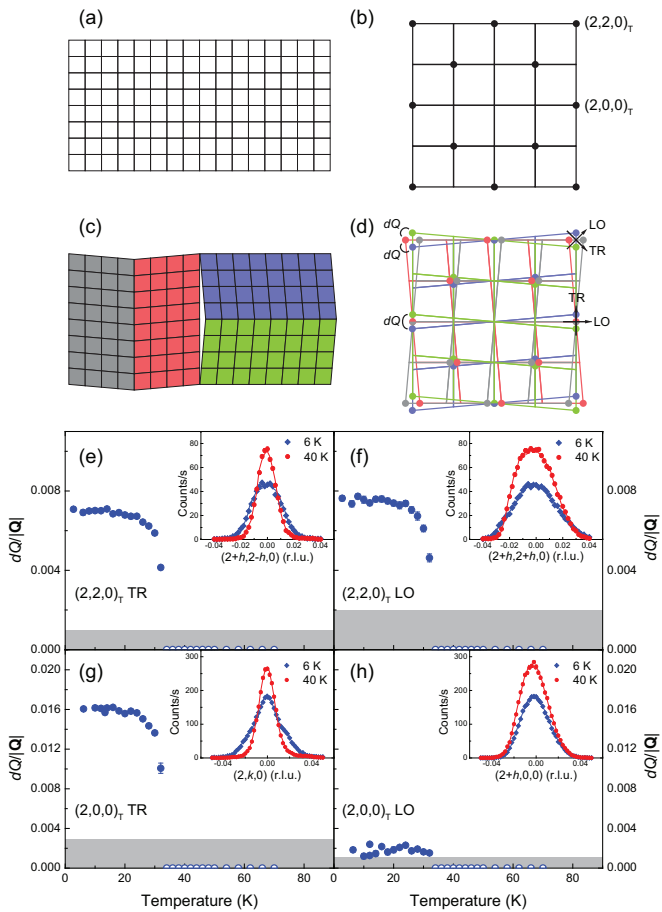}
	\caption{
		(Color online) Schematic of the (a) real and (b) reciprocal space lattices of BaFe$_2$As$_2$ in the tetragonal phase. Upon cooling below the tetragonal-to-orthorhombic transition temperature, four structural domains as shown in (c) form in BaFe$_2$As$_2$, resulting in splitting of Bragg peaks in reciprocal space shown in (d). 
		(a)-(d) are adapted from Ref. \cite{MATanatar2009}. 
		$dQ/|{\bf Q}|$ for KFe$_{0.8}$Ag$_{1.2}$Te$_2$ for (e) transverse and (f) longitudinal scans at $(2,2,0)_{\rm T}$. Similarly, for (g) transverse and (h) longitudinal scans at $(2,0,0)_{\rm T}$. For $T\geq35$ K, we do not observe broadening within our resolution (shaded gray area), therefore for these points $dQ/|{\bf Q}|$ are set to be zero (open symbols). The insets in (e)-(h) compare corresponding scans at 6 K and 40 K, the solid lines are fits to the data as described in the text. Error bars represent one standard deviation. 
	}
\end{figure} 

Our observation of intertwined stripe-type magnetic order and nematic order in KFe$_{0.8}$Ag$_{1.2}$Te$_2$ is different from observations of stripe-type magnetism in KFe$_{1.5}$Se$_2$ (K$_2$Fe$_3$Se$_4$) \cite{JZhao2012} and heavily Cu-doped NaFe$_{1-x}$Cu$_x$As \cite{YSong2016} and Ba(Fe$_{1-x}$Cu$_x$)$_2$As$_2$ \cite{WWang2017}. In the latter cases, there are no nematic orders that accompany the onset of magnetic orders. Since KFe$_{0.8}$Ag$_{1.2}$Te$_2$ is a semiconductor without itinerant electrons near the Fermi level \cite{HLei2011,RAng2013}, it provides an experimental benchmark to identify whether properties related the intertwined orders require contributions from itinerant electrons; properties intrinsic to the intertwined orders should be present in both iron pnictides and KFe$_{0.8}$Ag$_{1.2}$Te$_2$, whereas properties that require itinerant electrons would be present in iron pnictides but not in KFe$_{0.8}$Ag$_{1.2}$Te$_2$. The semiconducting nature of KFe$_{0.8}$Ag$_{1.2}$Te$_2$ further suggests local-moment magnetic interactions are responsible for its intertwined orders, different from the itinerant-electron scenarios proposed for the iron pnictides \cite{RMFernandes2012_1,RMFernandes2012_2}. Given the multi-orbital nature of Fe, orbital physics could also play an important role, although its manifestation in KFe$_{0.8}$Ag$_{1.2}$Te$_2$ would be quite different from that in the iron pnictides that exhibit split bands of $d_{xz}$ and $d_{yz}$ characters near the Fermi level \cite{MYi2011}.
We note that $T_{\rm N}\approx35$ K in KFe$_{0.8}$Ag$_{1.2}$Te$_2$ is considerably lower than $T_{\rm N}>500$ K in K$_{0.8}$Fe$_{1.6}$Se$_2$, most likely due to weak inter-block couplings that bottleneck the formation of long-range magnetic order. We expect that intra-block interactions should exhibit a much larger energy scale, with magnetic fluctuations that extend to high energies present around both ${\bf Q}=(0.5,0.5)_{\rm T}$ and ${\bf Q}=(0.5,-0.5)_{\rm T}$ in the paramagnetic state well above $T_{\rm N}$. 
The intertwined magnetic and nematic orders imply the presence of intense fluctuations of both, near the quantum critical point where the intertwined orders are suppressed. Given that superconductivity typically arises in the neighborhood of such putative quantum critical points in the iron pnictides \cite{TShibauchi2014}, suppressing the intertwined orders in KFe$_{0.8}$Ag$_{1.2}$Te$_2$ through doping or the application of pressure may lead to a novel superconducting state.

Our results also highlight the use of nonmagnetic elements to tune the physical properties of systems containing Fe-pnictogen/chalcogen planes, similar to iron-vacancy tuning in K$_z$Fe$_{2-y}$Se$_2$ \cite{WBao2011,JZhao2012}. Such tuning has been explored in $A$(Fe$_{1-x}B_x$)$_2Ch_2$ ($A=$ K, Na, Rb; $B=$ Li, Cu, Ag; $Ch=$ S, Se, Te) with $x\approx50\%$ \cite{DYuan2017,FSun2017}, 
and most such systems appear to exhibit a spin-glass ground state. 
Our crystal refinement results indicate that the ideal ratio of Fe and Ag is 2:3 in KFe$_{0.8}$Ag$_{1.2}$Te$_2$, this may explain why KFe$_{1.05}$Ag$_{0.88}$Te$_2$ exhibits a spin-glass ground state without long-range magnetic order \cite{HRyu2015,sg_note}. Similarly, the prevalence of a spin-glass ground state in $A$(Fe$_{1-x}B_x$)$_2Ch_2$ \cite{DYuan2017} may be due to most reported compounds being close to a 1:1 ratio of Fe and the non-magnetic $B$ element, rather than a ratio close to 2:3 as in KFe$_{0.8}$Ag$_{1.2}$Te$_2$. 
The availability of numerous similar systems \cite{DYuan2017} presents a unique opportunity to investigate the effects of magnetic dilution and chemical pressure on the physical properties of Fe-pnictogen/chalcogen planes, by tuning the ratio of Fe and the nonmagnetic $B$ element, and by replacing Ag with other nonmagnetic elements or Te with other chalcogens. 

In summary, we find that Fe atoms in KFe$_{0.8}$Ag$_{1.2}$Te$_2$ order into isolated $2\times2$ blocks, forming a $\sqrt{5}\times\sqrt{5}$ superstructure.
Below $T_{\rm N}\approx35$ K, magnetic moments within the Fe blocks form a stripe-type pattern accompanied by a nematic order that breaks fourfold rotational symmetry of the crystal structure, exhibiting a phenomenology similar to the iron pnictides. Since KFe$_{0.8}$Ag$_{1.2}$Te$_2$ is a semiconductor, these intertwined orders should be driven by local-moment magnetism originating from the minimal quartet of 4 Fe spins, and the underlying interactions
should be important for the physics of materials with Fe-pnictogen/chalcogen planes in general, including iron-based superconductors.  


The work at ORNL’s HFIR was sponsored by the Scientific User Facilities Division, Office of Science, Basic Energy Sciences (BES), US Department of Energy. The work at Lawrence Berkeley National Laboratory was supported by the Office of Science, Office of BES, Materials Sciences and Engineering Division, of the US DOE under contract No. DE-AC02-05-CH11231 within the Quantum Materials Program (KC2202). The work at Brookhaven National Laboratory was supported by the US DOE under contract No. DE-SC0012704. The identification of any commercial product or trade name does not imply endorsement or recommendation by the National Institute of Standards and Technology.


\begin{thebibliography}{}
\bibitem{sg_note} We note the spin-glass behavior in KFe$_{1.05}$Ag$_{0.88}$Te$_2$ may be due to disorder acting as a random field that pins the nematic order, resulting in a random-field nematic-glass. Due to the intertwined nature of the nematic and magnetic orders, the spin-glass behavior then follows from the nematic-glass in this scenario.
\end{thebibliography}
\end{document}